\documentclass[structabstract]{aa}  
\usepackage{graphicx}
\usepackage{txfonts}
\usepackage{lscape}
\usepackage{rotating}
\begin{document}
   \title{Infrared Emission from the Composite Grains: 
Effects of Inclusions and Porosities on the 10 and 18 $\mu m$ Features}


   \author{D. B. Vaidya
          \inst{1}
          \and
	  Ranjan Gupta 
          \inst{2}
          }

   \institute{ICCSIR, Ahmedabad-380009, India\\
              \email{deepak.vaidya@iccsir.org}
         \and
	     IUCAA, Post Bag 4, Ganeshkhind, Pune-411007, India\\
             \email{rag@iucaa.ernet.in}
             \thanks{Corresponding Author}
             }

   \date{Received April 30, 2010; accepted January 6, 2011} 
 
  \abstract
{}
{In this paper we study the effects of inclusions and porosities on the
emission properties of silicate grains and compare the model curves with the observed
infrared emission from circumstellar dust.}
{We calculate the absorption efficiency of the composite grain, made up
of a host silicate oblate spheroid and inclusions of ice/graphite/or voids, in the spectral
region 5.0-25.0$\mu m$. The absorption
efficiencies of the composite spheroidal oblate grains for three axial
ratios are computed using the discrete dipole approximation (DDA).
We study the absorption as a function of the volume fraction of
the inclusions and porosity. In particular, we study the variation
in the $10\mu m$ and $18\mu m$ emission features with the
volume fraction of the inclusions and porosities.
We then calculate the infrared fluxes
for these composite grains at several dust temperatures (T=200-350K)
and compare the model curves with
the average observed IRAS-LRS curve, obtained for 
circumstellar dust shells around oxygen rich M-type stars. 
The model curves are also compared with two other individual stars.}
{The results on the composite grains show variation in the absorption efficiencies 
with the variation in the inclusions and porosities. In particular, it is found that 
the wavelength of peak absorption at $10\mu m$, shifts towards longer wavelengths
with variation in the volume fraction of the inclusions of graphite. The spheroidal 
composite grains with axial ratio $\sim$ 1.33; volume fraction of f=0.1 and dust 
temperature between 210-340K, fit the observed infra-red emission from 
circumstellar dust reasonably well in the wavelength range 5-25$\mu m$.
The model flux ratio, R=Flux(18$\mu$)/Flux(10$\mu$), compares well with 
the observed ratio for the circumstellar dust.}
{The results on the composite grains clearly indicate that 
the silicate feature at 10$\mu m$ shifts with the volume fraction of
graphite inclusions. The feature does not shift with the porosity. Both the features
do not show any broadening with the inclusions or porosity. The absorption
efficiencies of the composite grains calculated using DDA and Effective Medium Approximation
(EMA) do not agree. The composite grain models presented in this study need to be
compared with the observed IR emission from the circumstellar dust around a few
more stars.}

   \titlerunning{IR emission from composite grains}

   \maketitle
%

\section{Introduction}

Circumstellar dust grains are more likely to be non-spherical and
inhomogeneous, viz. porous, fluffy and composites of many small grains glued
together, due to grain-grain collisions, dust-gas interactions and various
other processes. Since there is no exact theory to study the scattering properties
of these inhomogeneous grains, there is a need for formulating models of electromagnetic 
scattering by these grains. There are two widely used approximations to study the optical 
properties of composite grains viz.
effective medium approximation (EMA) and discrete dipole approximation (DDA).
We use DDA for calculating the absorption efficiencies of the composite grains. Mathis \&
Whiffen (1989) and Mathis (1996) have used EMA to calculate the absorption cross-section
for the composite grains containing silicate and amorphous carbon.
For details on EMA refer Bohren \& Huffman (1983) and for DDA refer Draine (1988).
For comparison of two methods see Bazell \& Dwek (1990);
Perrin \& Lamy (1990); Perrin \& Sivan (1990); Ossenkopf (1991); Wolff et. al. (1994)
and Iati et al. (2004).

In this paper we study the effects of inclusions and porosities on the absorption efficiencies
of the silicate grains in the wavelength range of 5--25 $\mu m$. In particular we study
the variation in the emission features at 10$\mu m$ and 18$\mu m$ with the volume fraction
of inclusions and porosities. We use these absorption efficiencies to compare the average
observed infrared emission curve obtained for the circumstellar dust around several
oxygen rich M-type stars (IRAS LRS catalogue of Olnon \& Raimond, 1986). We have also
compared the model curves with two individual stars.

Earlier studies by Henning \& Stognienko (1993) showed that composite oblate grains containing 
silicates and graphites did not show any changes in 10$\mu m$ and 18$\mu m$ features or the
ratio R=Flux(18$\mu$)/Flux(10$\mu$) with respect to the silicate grains. It must be noted
here that they used DDA for calculating absorption cross-section of the composite oblate
grains. O'Donnell (1994) also did not find any shift in the 10$\mu m$ or 18$\mu m$ features for the
grains containing silicates with the inclusions of carbons (glassy and amorphous). 
Min et. al. (2006, 2007) have used DDA to study the composite and aggregated silicates
and found that the 10$\mu m$ feature shifts to the shorter wavelengths. 
Jones (1988) found enhancement in the infrared absorption features at 
9.7$\mu m$ and 18$\mu m$ for porous silicate grains and hollow spheres.
In view of these studies
a detailed investigation of the 10$\mu m$ and 18$\mu m$ features using realistic grain models
is called for.

In section 2 we give the validity criteria for the DDA and the composite grain models.
In section 3 we present the results of our computations and compare
the model curves with the observed IR fluxes obtained by IRAS satellite.
Section 4 provides a detailed discussion on the comparison of our model/results with
available model/results from other workers.
The main conclusions of our study are given in section 5.


\section{Composite Grains and DDA}

We use the modified computer code (Dobbie, 1999) to generate the
composite grain models used in the present study.
We have studied composite grain models with a host silicate oblate
spheroid containing N= 9640, 25896 and 14440 dipoles, each carved out from
$32 \times 24 \times 24$, $48 \times 32 \times 32$
and  $48 \times 24 \times 24$ dipole sites, respectively;
sites outside the spheroid are set to be vacuum and sites inside are
assigned to be the host material.
It is to be noted that the composite spheroidal grain with
N=9640 has an axial ratio of 1.33, whereas N=25896 has the axial
ratio of 1.5, and N=14440 has the axial ratio of 2.0.
The volume fractions of the graphite inclusions used are
10\%, 20\% and 30\% (denoted as f=0.1, 0.2 and 0.3)
Details on the computer code and the corresponding modification
to the DDSCAT code (Draine \& Flatau 2003) are given in
Vaidya et al. (2001, 2007) and Gupta et al. (2006).
The modified code outputs a three-dimensional matrix specifying
the material type at each dipole site; the sites are either silicate,
graphite or vacuum. An illustrative example of a composite spheroidal oblate grain
with N=9640 dipoles, is shown in Figure 1. This figure also shows the inclusions 
embedded in the host oblate spheroid. 
Oblate spheroids were selected based on the numerous results of previous studies 
(Greenberg and Hong 1975; Henning and Stognienko 1993; O'Donnell 1994, Gupta et al 2005)
that showed that oblate spheroids better represent properties of circumstellar 
dust particles, specifically, this model provides a good fit to the observed 
polarization across the 10$\mu m$ feature (Lee and Draine 1985).

\begin{figure*}
\includegraphics[width=8.0cm]{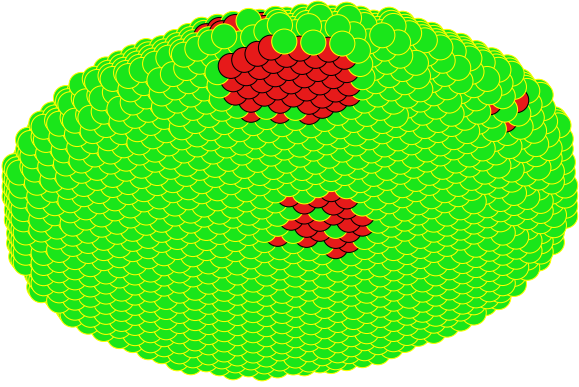}
\caption{A composite grain with a total of N=9640 dipoles, the inclusions are
embedded in the host oblate spheroid are shown (in red color).}
\end{figure*}

There are two validity criteria for DDA (see e.g. Wolff et al. 1994);
viz. (i) $\rm |m|kd \leq 1$, where m is the complex refractive index
of the material, k=$\rm \pi/\lambda$ is the wave number and
 d is the lattice dispersion spacing and
(ii) d should be small enough (N should be sufficiently large) to
describe the shape of the particle satisfactorily.
We have checked the validity criteria i.e. $\rm |m|kd \leq 1$,
for all the composite grain models with inclusions of ices, graphites and voids.
The $\rm |m|kd \leq 1$ varied from 0.041 at 5 $\mu m$ for N=9640 to 0.001 at
25 $\mu m$ for N=25896.

Table 1 shows the number of dipoles (N) for each grain model in the first column and also
the size of inclusion ('n' across the diameter of an inclusion e.g. 152 for N=9640
see Vaidya et al. 2001),
the remaining three columns show the number of inclusions and number
of dipoles per inclusion (in brackets) for the three volume fractions (f=0.1,
f-0.2 and f=0.3) respectively.

\begin{table*}
\begin{center}
\caption{Size of Inclusions, Number of inclusion (and number of dipoles per inclusion)}
\begin{tabular}{lccc}
\hline
\hline
Inclusions & Inclusion & Fractions &\\
\hline
N=9640 & f=0.1 & f=0.2 & f=0.3\\
AR=1.33 & 32/24/24 &&\\
\hline
n=16/12/12 & 1(1184)  & 2(1184) & \\
n=8/6/6 & 6(152) & 11(152) & 16(152)\\
n=4/3/3 & 38(16) & 76(16) & 114(16)\\
\hline
N=25896 & f=0.1 & f=0.2 & f=0.3\\
AR=1.50 & 48/32/32 &&\\
\hline
n=12/8/8 & 7(432)  & 13(432) & 19(432)\\
n=6/4/4 & 54(56) & 108(56) & 162(56)\\
n=3/2/2 & 216(8) & 432(8) & 648(8)\\
\hline
N=14440 & f=0.1 & f=0.2 & f=0.3\\
AR=2.00 & 48/24/24 &&\\
\hline
n=16/8/8 & 3(536) & 6(536) & 8(536)\\
n=12/6/6 & 6(224) & 11(224) & 16(224)\\
n=8/4/4 & 23(64) & 46(64) & 68(64)\\
n=6/3/3 & 38(24) & 76(24) & 114(24)\\
n=4/2/2 & 91(8) & 181(8) & 271(8)\\
\hline
\end{tabular}
\end{center}
\end{table*}

The complex refractive indices for silicates and graphite are obtained from 
Draine (1985, 1987) and that for ice is from (Irvine \& Pollack 1969).

As mentioned before the composite spheroidal grain
models with N=9640, 25896 and 14440 have the axial ratio 1.33,
1.5 and 2.0 respectively and if the semi-major axis and
semi-minor axis are denoted by x/2 and y/2 respectively, then
$\rm a^3=(x/2)(y/2)^2$, where 'a' is the
radius of the sphere whose volume is the same as that of
a spheroid.
In order to study randomly oriented spheroidal grains, it is
necessary to get the scattering properties of the composite
grains averaged over all of the possible orientations; in the
present study we use three values for each of the orientation
parameters ($\rm \beta, \theta and \phi$), i.e. averaging over 27 orientations,
which we find quite adequate (see Wolff et al. 1994).

\section{Results \& Discussion}

\subsection{Absorption Efficiency of Composite Spheroidal Grains}

Recently we have studied the effects of inclusions and porosities in the 
silicate grains on the infrared emission properties in the wavelength 
region 5-14 $\mu m$ (Vaidya and Gupta 2009).
In the present paper, we study the absorption properties of the
 composite spheroidal grains
with three axial ratios, viz. 1.33, 1.5 and 2.0, corresponding to
the grain models with N=9640, 25896 and 14440 respectively,
 for three volume fractions of inclusions; viz. 10\%, 20\% and 30\%,
in the extended wavelength region 5.0-25.0$\mu m$. The inclusions, selected
are graphites/ices/or voids. In this paper we particularly study the effects
of inclusions and porosity on the 10$\mu m$ and 18$\mu m$ features individually, 
as well as on the flux ratio R=Flux(18$\mu$)/Flux(10$\mu$).

Figures 2 (a-c) show the absorption efficiencies ($\rm Q_{abs}$) for the
 composite grains with the host silicate spheroids containing 9640,
 25896 and 14440 dipoles.
The three volume fractions, viz. 10\%, 20\% and 30\%, of ice
inclusions are also listed in the top (a) panel.
It is seen that there is no appreciable variation in the absorption efficiency
with the change in the volume fraction of inclusions in the wavelength region 5-8$\mu m$.
The variation in the absorption efficiency is clearly seen in the wavelength range
between 8-25$\mu m$ with peaks at 10$\mu m$ and 18$\mu m$.
It is also to be noted that there is no shift in the wavelength of the peak absorption.
In figures 2(d-f) variation of the absorption efficiency between 8 and 14$\mu m$
is highlighted. It is seen that the strength of both absorption peaks decrease with 
the increase of volume fraction of the inclusions. 

\begin{figure*}
\includegraphics[width=8.0cm]{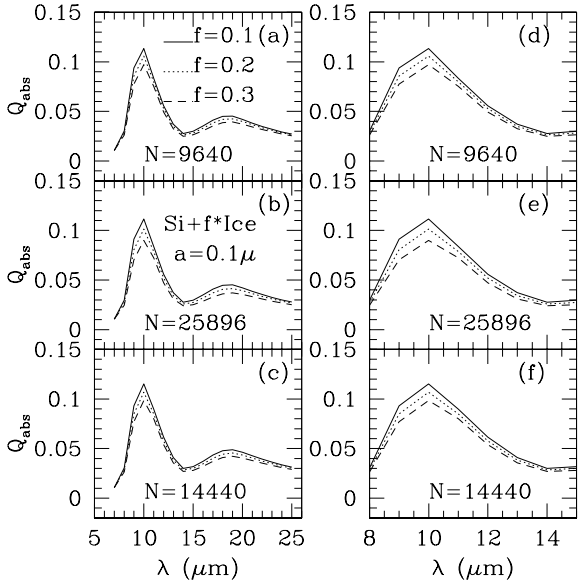}
\caption{Absorption Efficiencies for the composite grains with host silicate spheroids
and ices as inclusions for all three axial ratios N=9640 (AR=1.33);
N=25896 (AR=1.50) and N=14440 (AR=2.00). The 10$\mu$ feature is highlighted in the right side
panels (d-f).}
\end{figure*}

Figure 3 shows the absorption efficiencies for the composite grains with
the host silicate spheroids and graphite inclusions. It is seen in figures 3(d-f) that
the 10$\mu$ feature shifts towards shorter wavelengths as the
volume fraction of the graphite inclusions increases.
Ossenkopf et al. (1992) have studied the effects of inclusions of $Al_2 O_3, MgO, MgS$
and carbons (glassy and amorphous) in the silicate grains and they too
have found that the 10$\mu m$ absorption feature shifts shortwards.
O'Donnell (1994) did not find any shift in 10$\mu m$ feature for the silicate
grains with the inclusions of carbons. We did not find any shift
in the absorption feature at 18$\mu m$ with the change in the volume fraction
of the graphite inclusions. Ossenkopf et al. (1992) and O'Donnell (1994) also did
not find any variation in the 18$\mu m$ feature with the inclusions.
Henning and Stognienko (1993) have used composite oblate spheroid grains containing
silicates and graphites and found no significant shift in the 10$\mu m$ or 18$\mu m$ features.

Results in Figures 3(a-f) also indicate that absorption efficiency does not
vary with the shape of the grains (axial ration AR=1.33, 1.50, 2.00).

We also checked the absorption efficiencies of the composite grains for
several inclusion sizes (n) at a constant volume fraction (see Table 1 e.g for N=9640
the array 8/6/6 indicates the size of the inclusions). We did not find any
significant change in the absorption efficiency or any shift in the absorption features
(Vaidya et al. 2001).

\begin{figure*}
\includegraphics[width=8.0cm]{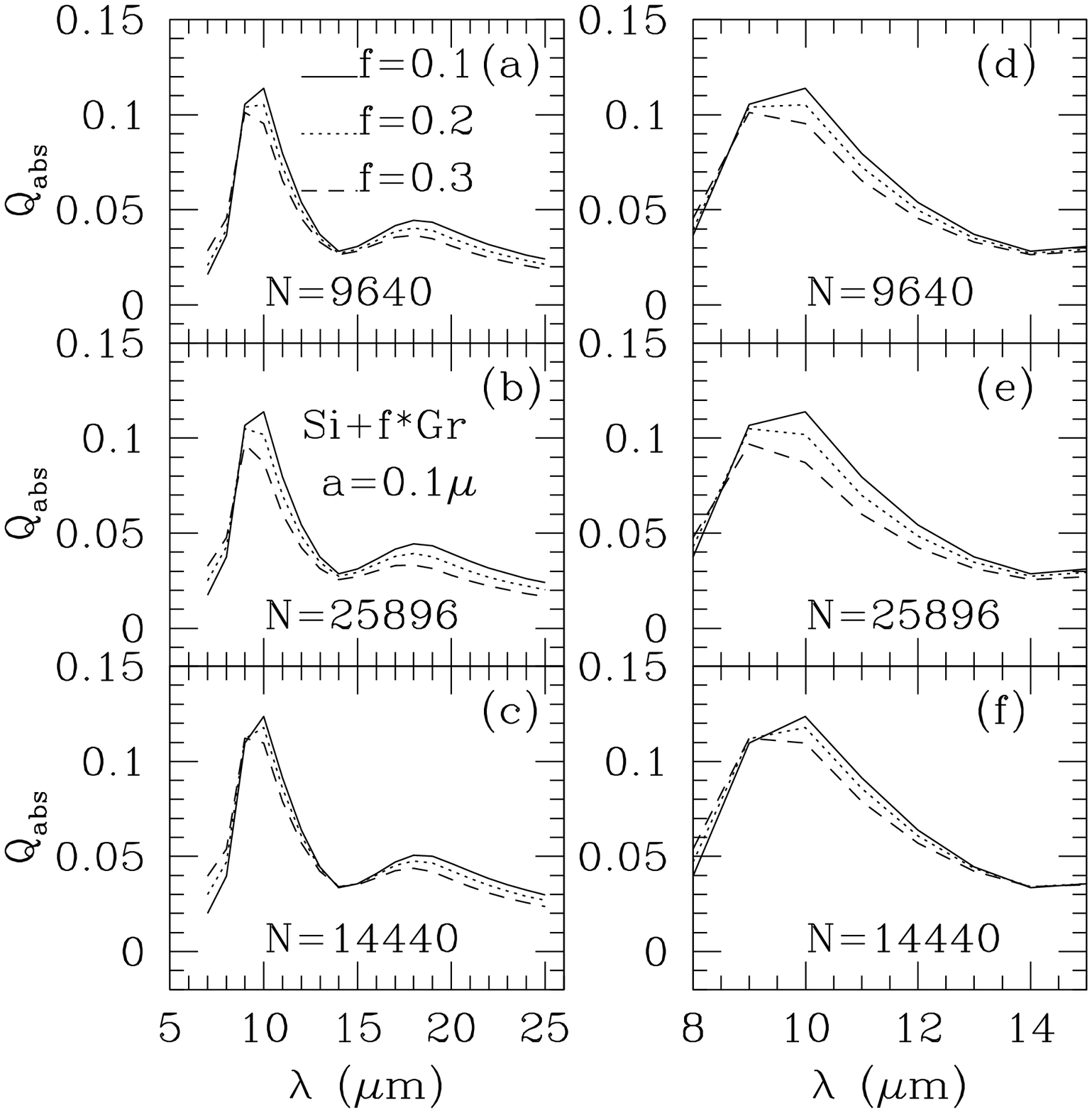}
\caption{Absorption Efficiencies for the composite grains with host silicate spheroids
and graphites as inclusions for all three axial ratios N=9640 (AR=1.33);
N=25896 (AR=1.50) and N=14440 (AR=2.00). The 10$\mu$ feature is highlighted in the right side
panels (d-f).}
\end{figure*}

We have compared our results on the absorption efficiencies
of the composite grains obtained using the DDA with the results obtained
using the EMA-T-matrix based calculations. The results with EMA are
displayed in Figure 4. For these calculations, the optical constants
were obtained using the Maxwell-Garnet mixing rule (i.e.
effective medium theory, see Bohren and Huffman, 1983). Description
of the T-matrix method/code is given by Mishchenko et al. (2002).

\begin{figure*}
\includegraphics[width=8.0cm]{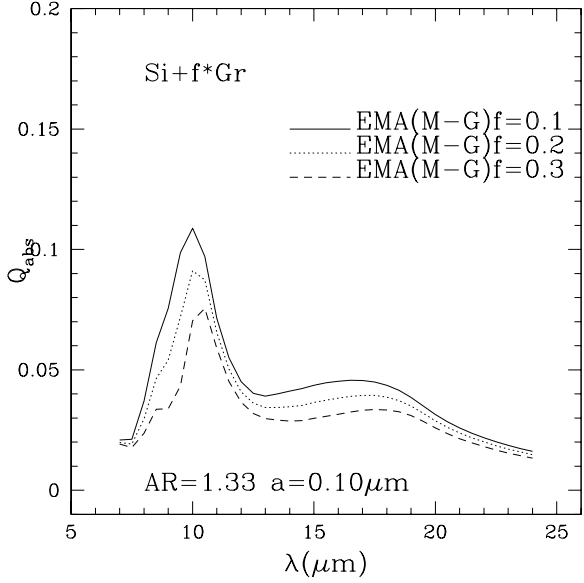}
\caption{EMA(M-G) calculations with AR=1.33 and three volume fractions.}
\end{figure*}

In Figure 5 we show the ratio Q(EMA)/Q(DDA) to compare the results obtained using both
the methods. It is seen that the absorption curves obtained using the
EMA-T matrix calculations, deviate from the absorption curves obtained using
the DDA, as the volume fraction of inclusions increases.
The results based on the EMA-T-matrix calculations and DDA results do not agree
because the EMA does not take into account the inhomogeneities within the
grain (viz. internal structure, surface roughness, voids; see Wolff et al. 1994, 1998)
and material interfaces and shapes are smeared out into a homogeneous 'average
mixture' (Saija et al. 2001). However, it would still be very useful and desirable
to compare the DDA results for the composite grains with those computed by other
EMA/Mie type/T matrix techniques in order to examine the applicability
of several mixing rules e.g. see Wolff et al. 1998, Voshchinnikov and Mathis
1999, Chylek et al. 2000, Voshchinnikov et al. 2005, 2006. The application of DDA,
poses a computational challenge, particularly for the large values of
the size parameter X ($\rm=2\pi a/\lambda > 20$ ) and the complex refractive index m of
of the grain material would require large number of dipoles and that in
turn would require considerable computer memory and cpu time (see e.g. Saija et al.
2001, Voshchinnikov et al. 2006).

\begin{figure*}
\includegraphics[width=8.0cm]{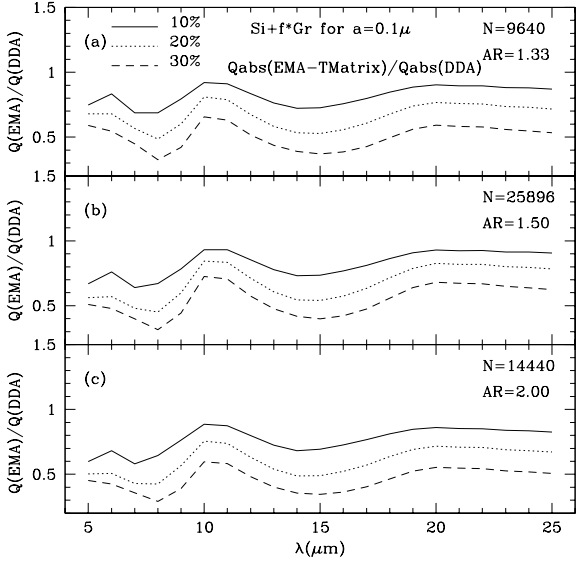}
\caption{Ratio for absorption efficiency using DDA and EMA.}
\end{figure*}

We have also calculated the absorption efficiencies of the porous grains.
Figure 6 shows the absorption efficiencies of the composite grains with
the host silicate spheroids and voids as inclusions.
It is seen that as the porosity increases i.e., as the volume fraction 'f' of the voids
increases, the peak strength decreases.
However, we did not find any shift in the 10$\mu m$ and 18$\mu m$ features with porosity.
Henning and Stognienko (1993) also did not find any change in the 10$\mu m$ or 18$\mu m$ feature
for the porous silicate grains.
Greenberg and Hage (1990) have shown the change in the feature strength and its
shape with the porosity of the grain.
Voshchinnikov et al. (2006) and Voshchinnikov \& Henning (2008) have used layered
sphered model to study the effect of porosity on the 10$\mu m$ feature and they found
that the peak strength decreases and the feature broadens with the porosity.
Recently, Li et al (2008) have used the porous grains to model the 10$\mu m$ feature
in the AGN and they found shift in the 10$\mu m$ absorption peak towards longer
wavelength. Min et al. (2007), have successfully used DDA to study the 10$\mu m$ silicate feature
of fractal porous grains and explain the interstellar extinction in various lines of sight.

\begin{figure*}
\includegraphics[width=8.0cm]{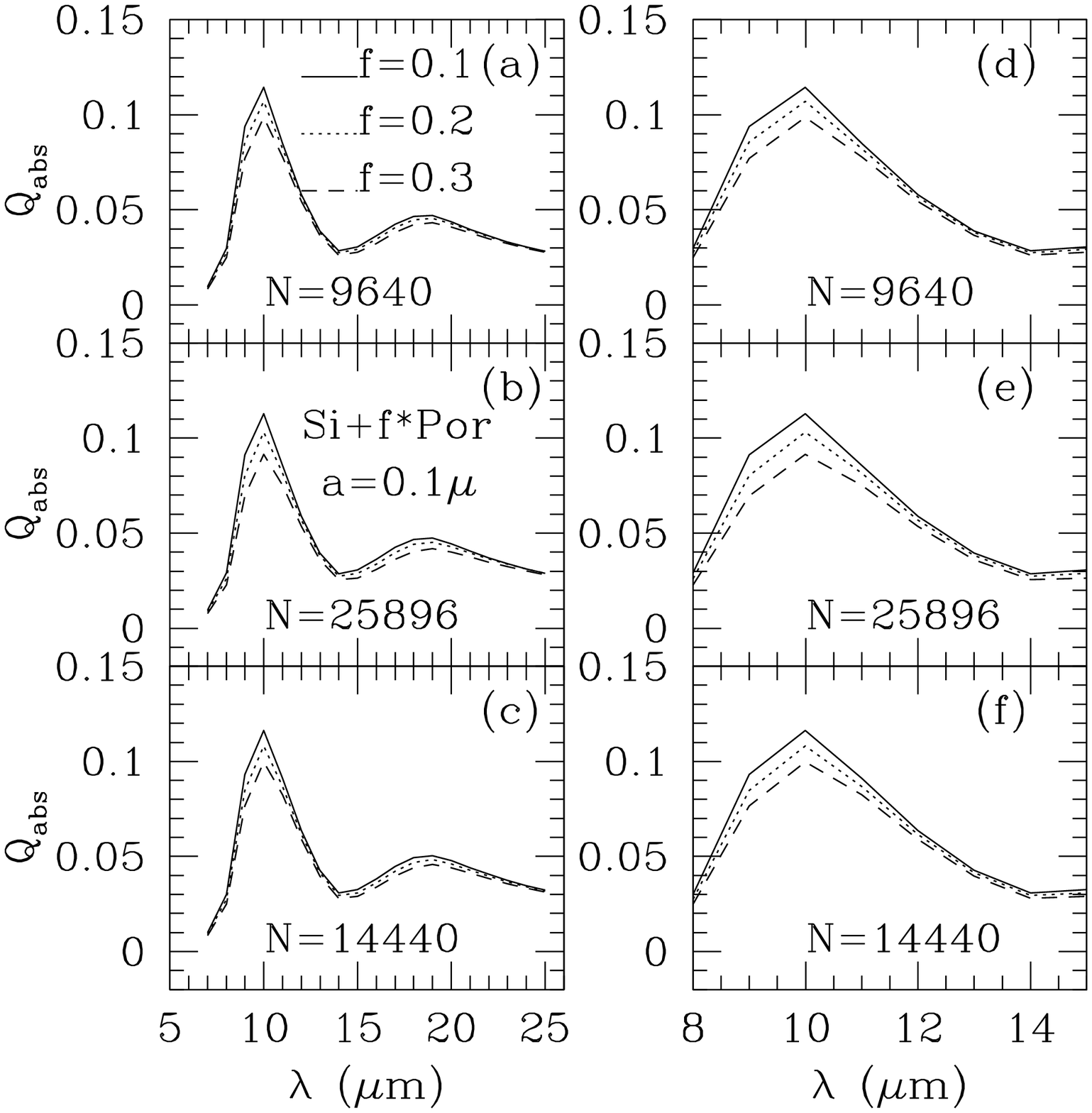}
\caption{Absorption Efficiencies for the composite grains with host silicate spheroids
and voids (vacuum) as inclusions for all three axial ratios N=9640 (AR=1.33);
N=25896 (AR=1.50) and N=14440 (AR=2.00). The 10$\mu$ feature is highlighted
in the right side panels (d-f).}
\end{figure*}

Figures 7, 8 and 9 show the variation of absorption efficiencies with the grain sizes
for the composite grains viz. a=0.05, 0.1, 0.5 and 1.0$\mu$, with the fraction of
inclusion of ices, graphites and voids respectively.
It is seen that for the small sizes viz. a=0.05 and 0.1$\mu$, the variation in the
absorption efficiency with the change in the volume fraction of inclusions is not
appreciable, whereas for the larger grains (a=0.5 and 1.0$\mu$) the effect is clearly
seen i.e. absorption efficiency decreases with the increasing fraction of inclusions.
In the Figure 8 we show the Q$_{abs}$ for the silicate
grain (i.e. volume fraction f=0.00). It is seen that the absorption is higher than 
that for composite grains.

\begin{figure*}
\includegraphics[width=8.0cm]{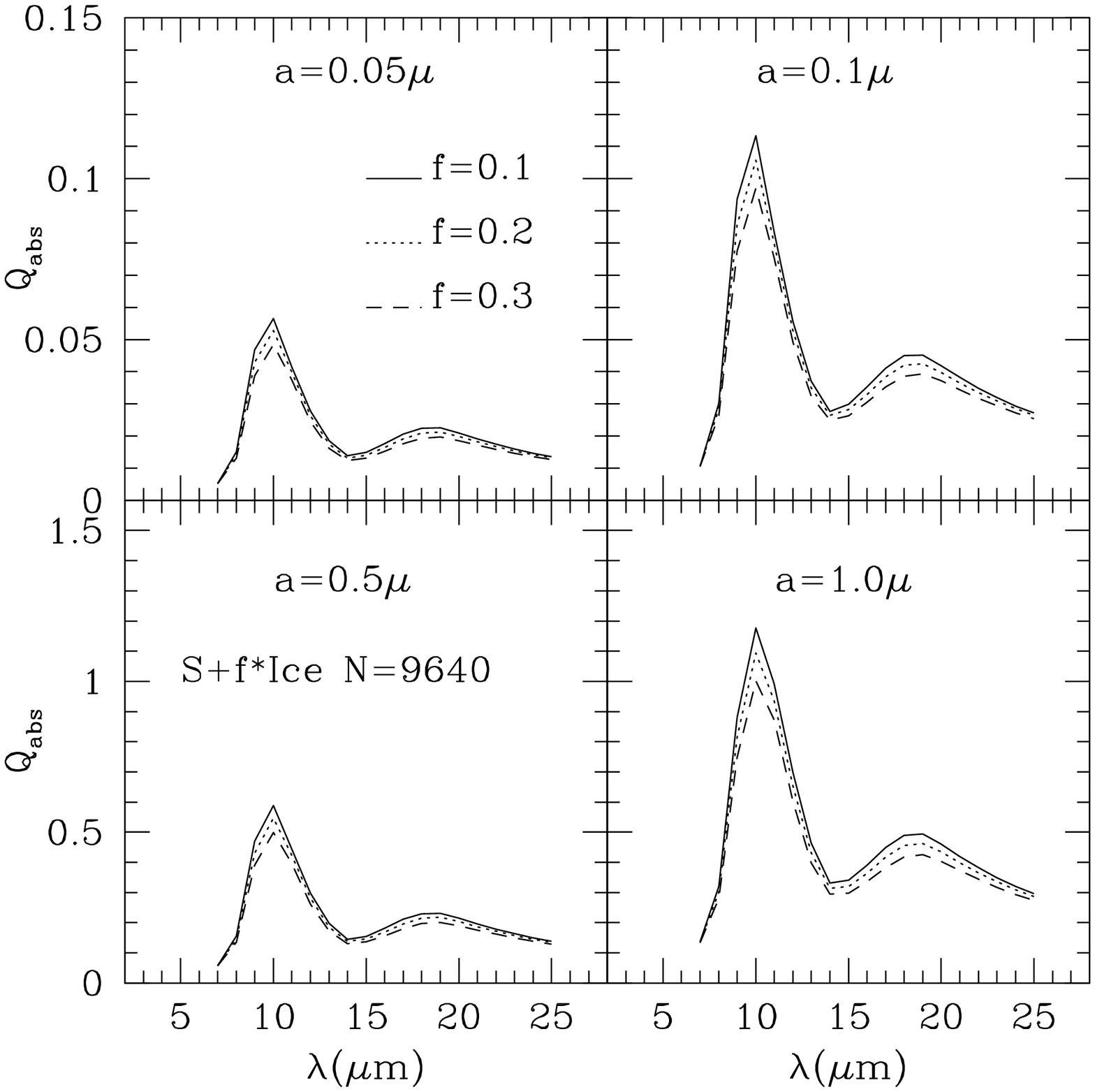}
\caption{Variation of Absorption Efficiencies with composite grains sizes.
Host silicate spheroids contain dipoles N=9640 and ices as inclusions}
\end{figure*}

\begin{figure*}
\includegraphics[width=8.0cm]{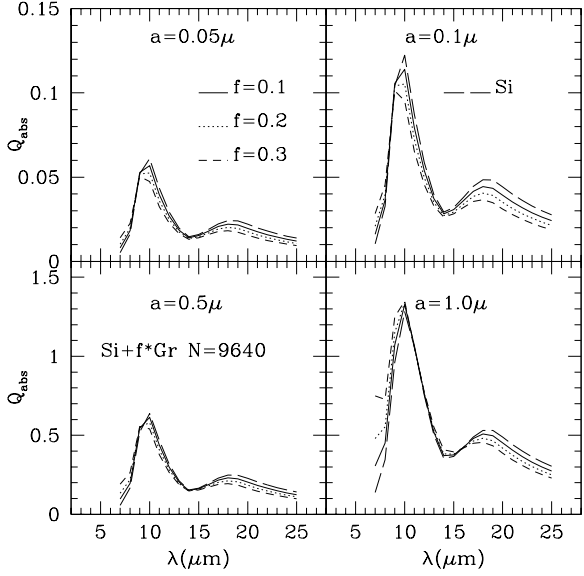}
\caption{Variation of Absorption Efficiencies with grains sizes.
Host silicate spheroids contain dipoles, N=9640 and graphites as inclusions.
Also shown is the Q$_{abs}$ for the silicate grain (f=0.0) for all the sizes.}
\end{figure*}

\begin{figure*}
\includegraphics[width=8.0cm]{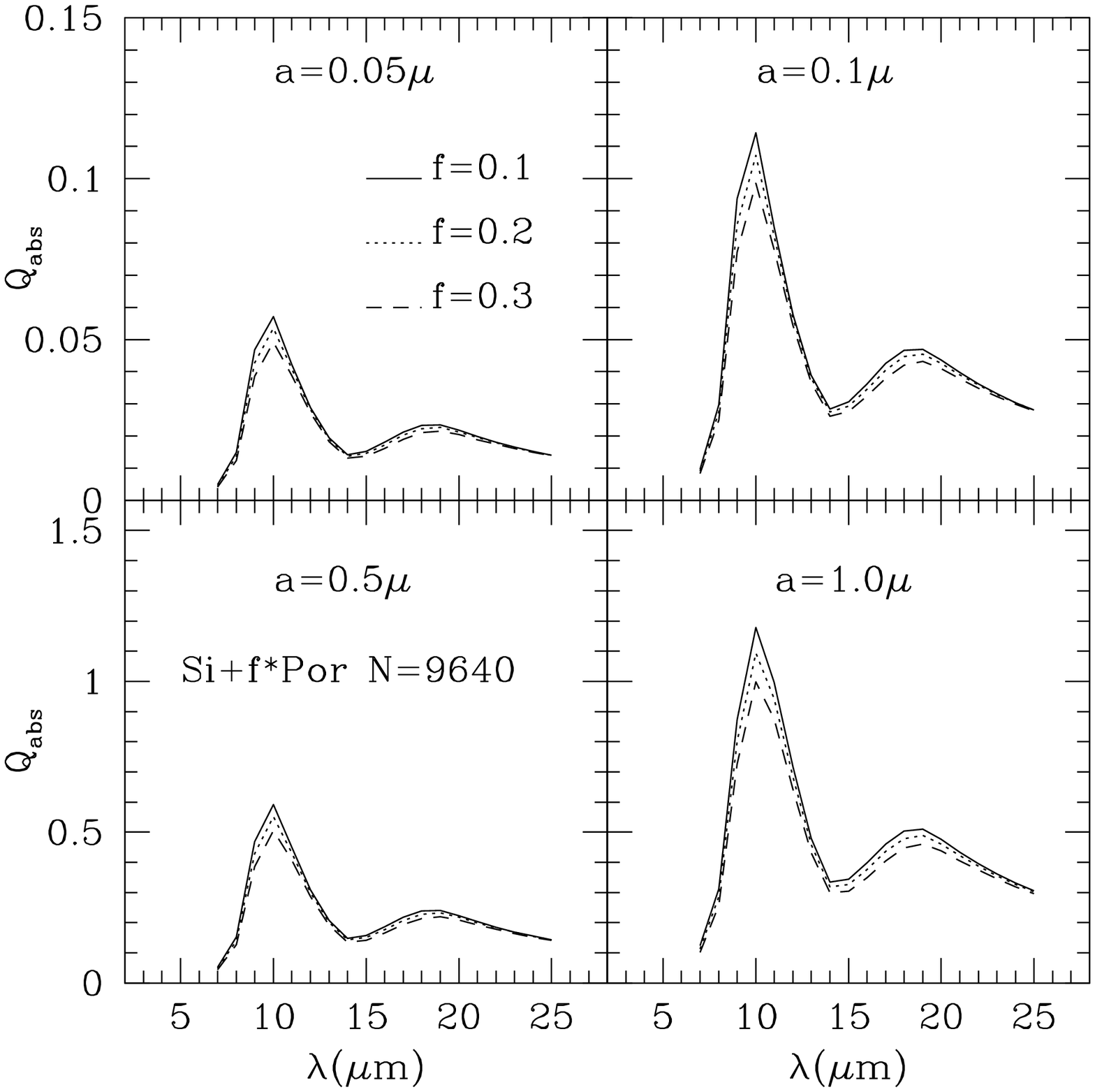}
\caption{Variation of Absorption Efficiencies with composite grains sizes.
Host silicate spheroids contain dipoles N=9640 and voids (vacuum) as inclusions}
\end{figure*}

All these results on the composite grain models show variation in the absorption
efficiencies with the variation of the volume fraction of the inclusions and porosities. 
These results
also show that the peak absorption wavelength at 10$\mu m$ shifts with the graphite
inclusions. These composite grain models do not show any shift in the
absorption peak at 18$\mu m$ with the change in the volume fraction of the inclusions.
Our results on the composite grain models do not show any broadening of the 10$\mu m$ or 18$\mu m$
feature.

\subsection{Infrared Emission from Circumstellar Dust:
Silicate features at 10$\mu m$ and 18$\mu m$}

In general, stars which have evolved off the main sequence and
which have entered the giant phase of their evolution are a major
source of dust grains in the galactic interstellar medium. Such
stars have oxygen overabundant relative to carbon and therefore
produce silicate dust and show the strong feature
at 10$\mu m$.  This is ascribed to the Si-O stretching mode in
some form of silicate material, e.g. olivine. These materials
also show a much broader and weaker feature at 18$\mu m$,
resulting from the O-Si-O bending mode (Little-Marenin and Little, 1990).
Using the absorption efficiencies of the composite grains and a power law
MRN dust grain size distribution (Mathis et al., 1977), we calculate
the infrared flux F$_{\lambda}$ at various dust temperatures and compare the observed
IRAS-LRS curves with the calculated infrared fluxes, F$_{\lambda}$ for the
composite grain models. The flux F$_{\lambda}$ is calculated using the
relation F$_{\lambda}=Q_{abs}.B_{\lambda}(T)$ at dust temperature T
in K and B$_{\lambda}$ as the Planck's function.
This is valid only if the silicate emission region is optically thin (Simpson 1991;
Ossenkopf et al., 1992 and Li et al. 2008).
Figure 10 shows the IR fluxes with various dust temperatures
(T=200-350K) for the composite grains with N=9640, and inclusions
of graphites with f=0.1 and MRN (Mathis et al., 1977) grain size
distribution a=0.005-0.250$\mu$. We checked the grain models with
larger grain size distribution (a=0.1-1.0 $\mu$) and found that it did not
match the observed curve satisfactorily -- the fit was very poor.

\begin{figure*}
\includegraphics[width=8.0cm]{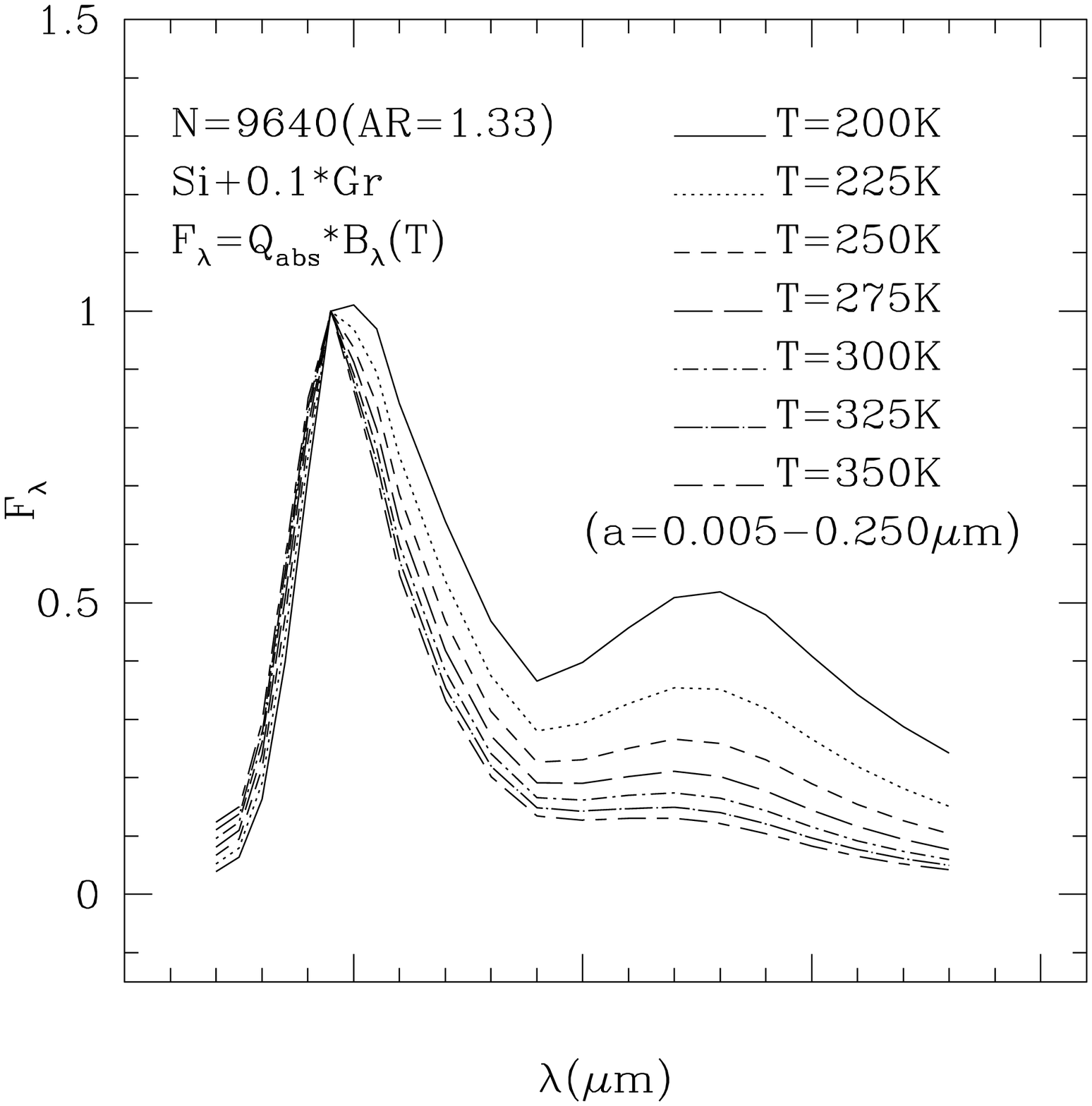}
\caption{Infrared Flux at various temperatures,
for the composite grains with graphites as inclusions}
\end{figure*}

Table 2 shows the best fit $\chi^{2}$ minimized values  and corresponding
temperatures for all the composite grain models with silicate host and graphite as
inclusions. For details on $\chi^{2}$ minimization please refer Vaidya \& Gupta (1997 and 1999).

\begin{table*}
\begin{center}
\caption{Minimum $\chi^{2}$ values and corresponding temperatures
(K in brackets)
for Si+f*Gr composite grain models fitting with average IRAS-LRS observed IR flux
and the two stars IRAS 16340-4634 and IRAS 17315-3414 and three volume fractions
of inclusions viz. f=0.1, 0.2 and 0.3.}
\begin{tabular}{lccc}
\hline\hline
Average Observed &&&\\
IRAS-LRS flux &&&\\
(Si+f*Gr) &&&\\
\hline
Inclusion fraction (f) & N=9640 & 25896 & 14440\\
\hline
0.1 & 0.00249(270) & 0.00260(270) & 0.00391(290)\\
0.2 & 0.00273(265) & 0.00332(260) & 0.00450(280)\\
0.3 & 0.00410(260) & 0.00573(260) & 0.00596(275)\\
\hline
IRAS-LRS &&&\\
16340-4634 flux &&&\\
(Si+f*Gr) &&&\\
\hline
0.1 & 0.00256(210) & 0.00156(210) & 0.00241(220)\\
0.2 & 0.00191(210) & 0.00165(210) & 0.00164(220)\\
0.3 & 0.00148(210) & 0.00156(210) & 0.00151(215)\\
\hline
IRAS-LRS &&&\\
17315-3414 flux &&&\\
(Si+f*Gr) &&&\\
\hline
0.1 & 0.00814(230) & 0.00846(235) & 0.00798(245)\\
0.2 & 0.00972(230) & 0.01112(230) & 0.00985(240)\\
0.3 & 0.01251(230) & 0.01474(230) & 0.01256(240)\\
\hline
\end{tabular}
\end{center}
\end{table*}

Table 3 shows the best fit $\chi^{2}$ minimized values  and corresponding
temperatures for all the composite grain models with silicate host and voids
(porous) as inclusions.

\begin{table*}
\begin{center}
\caption{Minimum $\chi^{2}$ values and corresponding temperatures
(K in brackets)
for and Si+f*Por composite grain models fitting with average IRAS-LRS observed IR flux
and the two stars IRAS 16340-4634 and IRAS 17315-3414 and three volume fractions
of inclusions viz. f=0.1, 0.2 and 0.3.}
\begin{tabular}{lccc}
\hline\hline
Average Observed &&&\\
IRAS-LRS flux &&&\\
(Si+f*Por) &&&\\
\hline
Inclusion fraction (f) & N=9640 & 25896 & 14440\\
\hline
0.1 & 0.00387(290) & 0.00442(300) & 0.00526(310)\\
0.2 & 0.00472(305) & 0.00575(320) & 0.00613(325)\\
0.3 & 0.00570(315) & 0.00708(340) & 0.00713(340)\\
\hline
IRAS-LRS &&&\\
16340-4634 flux &&&\\
(Si+f*Por) &&&\\
\hline
0.1 & 0.00400(220) & 0.00427(220) & 0.00450(225)\\
0.2 & 0.00451(220) & 0.00505(225) & 0.00517(230)\\
0.3 & 0.00543(220) & 0.00599(230) & 0.00591(230)\\
\hline
IRAS-LRS &&&\\
17315-3414 flux &&&\\
(Si+f*Por) &&&\\
\hline
0.1 & 0.00644(245) & 0.00629(245) & 0.00660(250)\\
0.2 & 0.00615(245) & 0.00599(250) & 0.00655(255)\\
0.3 & 0.00609(250) & 0.00625(260) & 0.00661(260)\\
\hline
\end{tabular}
\end{center}
\end{table*}

Figure 11(a) shows the average IRAS-LRS observed curve (Whittet, 2003)
and its comparison with the $\chi^{2}$ minimized best fit model N=9640
and f=0.1 graphite inclusions and a temperature
of T=270K. The Figure 11(b) and (c) show the observed
IRAS-LRS spectra of two typical stars
which have strong silicate feature at 10$\mu$ (IRAS class 6 as defined by Volk -- see
Olnon and Raimond, 1986; Gupta et al. 2004) viz. IRAS 16340-4634 and IRAS 17315-3414.
These two IRAS objects have been taken from the large set of 2000 IRAS spectra which
were classified into 17 classes bye eye (see Gupta et al. 2004) and have least
problems with noise or spectral peculiarities.
The first star IRAS 16340-4634 fits best with the $\chi^{2}$ minimized model N=25896,
and f=0.3 graphite inclusions and a temperature of T=210K.
The second star IRAS 17315-3414 fits best with the $\chi^{2}$ minimized
model N=14440, and f=0.1 graphites inclusions and a temperature of T=245K.

Figures 12 (a), (b) and (c) are for the same IRAS-LRS average observed curve
and the two IRAS stars respectively but best fitted to silicate host and voids (porous)
as inclusions.

The results of model fits to the corresponding temperatures in Figures 11 and 12 lie within
a range of 210-290K which essentially indicates the expected range of dust temperatures
in the circumstellar disks. One needs to compare the models with a larger set of
observed spectra to make more definitive estimates of dust temperatures for individual
IRAS and other sources (Henning \& Stognienko, 1993).

\begin{figure*}
\includegraphics[width=8.0cm]{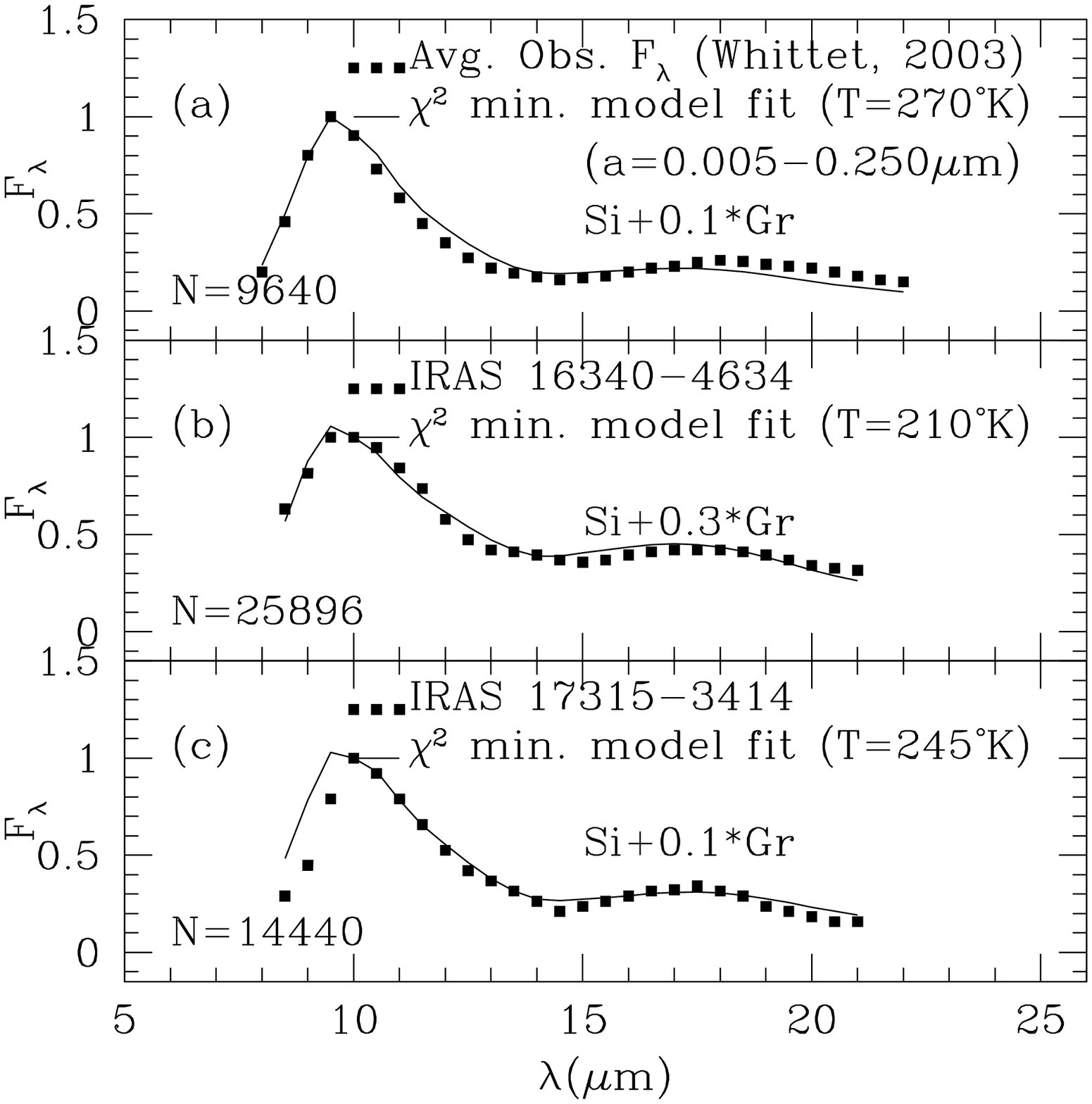}
\caption{Best fit $\chi^{2}$ minimized composite grain models (silicates with graphite
inclusions) plotted
with the average observed infrared flux for IRAS-LRS curve and the two stars
the IRAS 16340-4634 and IRAS 17315-3414.}
\end{figure*}

\begin{figure*}
\includegraphics[width=8.0cm]{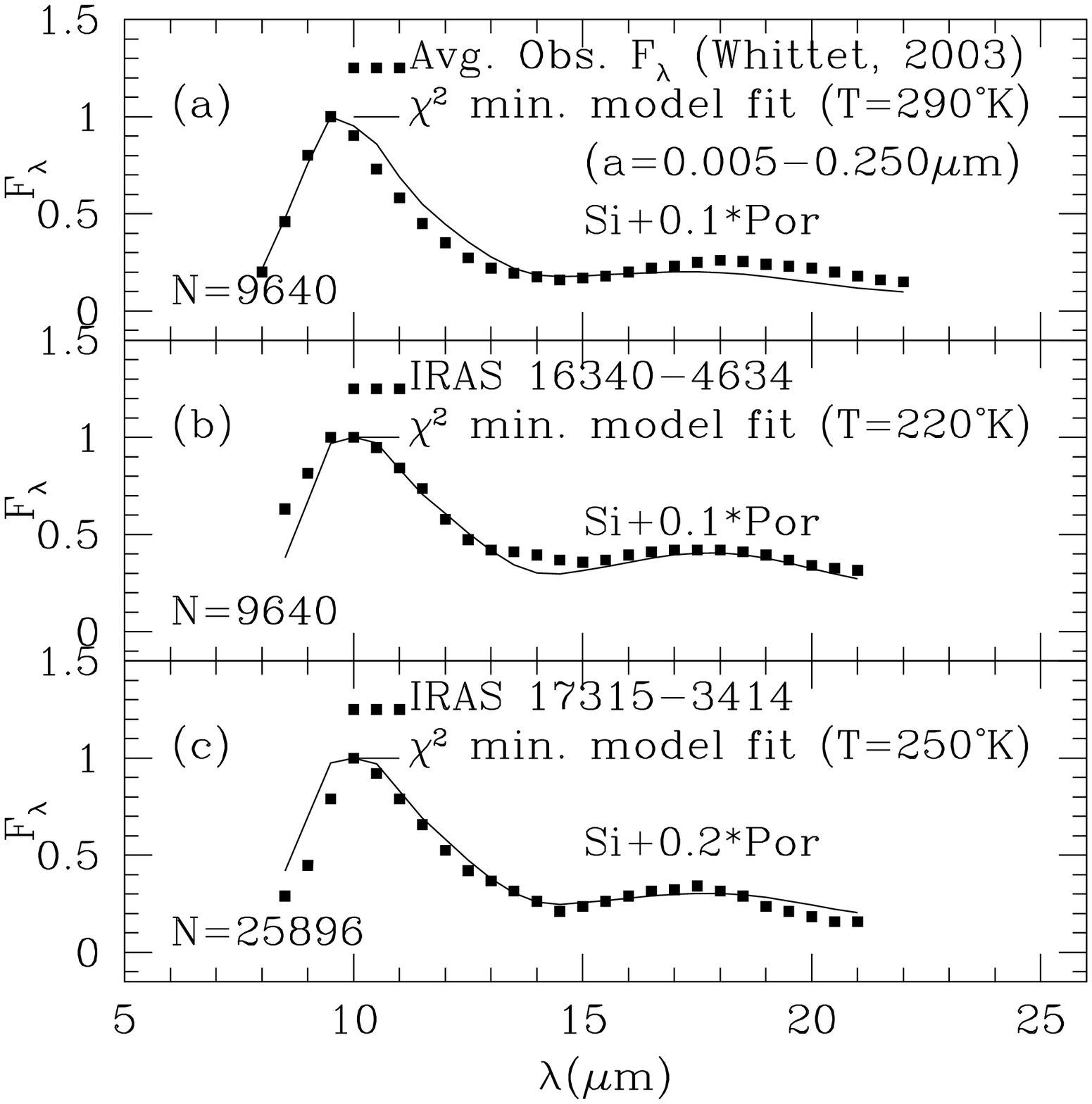}
\caption{Best fit $\chi^{2}$ minimized composite grain models (silicates with porous
inclusions) plotted
with the average observed infrared flux for IRAS-LRS curve and the two stars
the IRAS 16340-4634 and IRAS 17315-3414.}
\end{figure*}

It is to be noted that for the comparison with observed curves, we have not considered
composite grain models with ice as inclusions.
Ice is expected to condense in an O-rich stellar atmosphere if sufficiently low
temperature environments exist. Such conditions arise if the atmosphere is optically
thick (Whittet, 2003).
Hoogzaad et al. (2002) have used core-mantle grain model with core silicate
and ice as mantle to model the IR emission from the AGB star HD161796 and found that
the core-mantle dust model with ice as mantle has temperature in the range
of 50-75K.

\subsection{Flux Ratio R=Flux(18$\mu$)/Flux(10$\mu$)}

We have also studied the effect of inclusions and porosity in the silicate grain
on the flux ratio R=Flux(18$\mu$)/Flux(10$\mu$). Table 4 shows the ratio R for
composite grain models with graphite (Si+f*Gr) and voids (Si+f*Por) as inclusions.
It is seen from this table that in general for both the models the ratio R decreases
with the temperature; the ratio varies from $\sim$ 0.6 at T=200K to
$\sim$ 0.2 at T=300K. It is also to be noted from this table that
for the composite grain models with graphite as inclusions, the ratio R decreases
with the volume fraction of the inclusions, whereas for the models with the voids
(i.e. porous grains), the ratio R increases with the volume fraction of voids.
These results show that R increases with the porosity and thus clearly indicate that
both the inclusions and porosities within the grains modify the emission features in 
the silicate grains. Henning and Stognienko (1993) did not find any variation in the ratio
with the increase of the porosity. They also did not find any variation in the R with
the inclusions of graphite.

\begin{table*}
\begin{center}
\caption{The ratio of silicate features R=Flux(18$\mu$)/Flux(10$\mu$) for composite
grain models.}
\begin{tabular}{lccccccccc}
\hline\hline
Si+f*Gr  & 9640 &&&25896&&&14440&\\
\hline
T(K) & 0.1 & 0.2 & 0.3 & 0.1 & 0.2 & 0.3 & 0.1 & 0.2 & 0.3\\
\hline
200 & 0.504 & 0.485 & 0.471 & 0.503 & 0.478 & 0.463 & 0.551 & 0.528 & 0.504\\
225 & 0.339 & 0.325 & 0.316 & 0.338 & 0.321 & 0.310 & 0.370 & 0.354 & 0.338\\
250 & 0.247 & 0.237 & 0.231 & 0.247 & 0.234 & 0.227 & 0.270 & 0.259 & 0.247\\
275 & 0.192 & 0.184 & 0.179 & 0.191 & 0.181 & 0.176 & 0.209 & 0.200 & 0.191\\
300 & 0.156 & 0.149 & 0.145 & 0.155 & 0.147 & 0.143 & 0.170 & 0.163 & 0.155\\
\hline
Si+f*Por & 9640 &&&25896&&&14440&\\
\hline
T(K) & 0.1 & 0.2 & 0.3 & 0.1 & 0.2 & 0.3 & 0.1 & 0.2 & 0.3\\
\hline
200 & 0.567 & 0.592 & 0.619 & 0.584 & 0.620 & 0.655 & 0.607 & 0.631 & 0.656\\
225 & 0.381 & 0.398 & 0.416 & 0.392 & 0.416 & 0.439 & 0.408 & 0.423 & 0.440\\
250 & 0.278 & 0.290 & 0.303 & 0.286 & 0.304 & 0.321 & 0.298 & 0.309 & 0.321\\
275 & 0.215 & 0.225 & 0.235 & 0.220 & 0.236 & 0.249 & 0.231 & 0.240 & 0.249\\
300 & 0.175 & 0.183 & 0.191 & 0.180 & 0.191 & 0.202 & 0.187 & 0.194 & 0.202\\
\hline
\end{tabular}
\end{center}
\end{table*}

Table 5 shows the ratio of silicate features R=Flux(18$\mu$)/Flux(10$\mu$) for
the average IRAS-LRS observed curve and the two stars mentioned above and for
the best fit corresponding models (from Figures 11 and 12).
It is seen from Table 5 that in general the model ratio R=Flux(18$\mu$)/Flux(10$\mu$) 
for the average
observed curve (Whittet, 2003) is lower than that obtained for the two stars viz.
IRAS 16340-4634 and IRAS 17315-3414.
The model ratio, 0.383 for the star IRAS 16340-4634 is comparable with that derived 
for the circumstellar dust i.e. 0.394 (Simpson 1991 and Ossenkopf et al. 1992).
The low value of the ratio derived for the average observed circumstellar features
may be due to O-deficient silicates as noted by Little-Marenin and Little (1990)
and Ossenkopf et al. (1992). Ossenkopf et al. (1992) have noted that observationally
determined flux ratio R for circumstellar dust varies from 0.3 to 0.6.
However, it must be noted that the variation in R is not very significant if the range
of R, 10$\mu m$ and 18$\mu m$ features is considered. We need to compare the composite grain
model with a larger sample of stars to interpret the R for various stellar environments
as has been noted by Simpson (1991) and Henning \& Stognienko (1993).

\begin{table*}
\begin{center}
\caption{The Ratio R=Flux(18$\mu$)/Flux(10$\mu$) for the models and observed
silicate features.}
\begin{tabular}{lccc}
\hline\hline
Object/Star & R(Model Si+f*Gr) & R(Model Si+f*Por) & R(Observed)\\
\hline
Average Observed IRAS-LRS flux & 0.201 & 0.189 & 0.255\\
IRAS 16340-4634 & 0.383 & 0.379 & 0.394\\
IRAS 17315-3414 & 0.276 & 0.283 & 0.236\\
\hline
\end{tabular}
\end{center}
\end{table*}

\section{Comparison of our model/results with available model/results from
other workers}

In this section we have made a detailed comparison of the results of our present
work with various other published model/results on the silicate IR emission features which
are elaborated in the Tables 6 and 7 (Please note that for the discussion in these tables,
the peak position of the
10$\mu m$ feature lies in the interval 9.5-10.2$\mu m$ and the FWHM in the range
of 1.8-3.2$\mu m$, as derived by several authors (see Ossenkopf et al. 1992)).

It must be noted here that the list of composite and porous grain models given in Table 6 \& 7
are not exhaustive; we have included, particularly, the references/models which show 
variation in the 10$\mu m$ and 18$\mu m$ silicate features with volume fraction of inclusions 
or porosity.

These tables show that only three authors viz. Vaidya and Gupta (2010);
Henning \& Stognienko, (1993) and Min et al. (2007); have used DDA for
modeling the composite grains. Henning and Stognienko (1993)
did not find any significant variation either in 10$\mu m$, 18$\mu m$ features
or in the ratio R=Flux(18$\mu$)/Flux(10$\mu$) with porosity or inclusions. They have also
noted that the porous and composite grains are not the carriers of 
AFGL or BN objects. Min et al. (2007) have used a statistical ensemble
of simple particle shapes to represent irregularly shaped particles and 
the models fit the interstellar extinction profile in the spectral range of 5-25$\mu m$.
As mentioned earlier the EMA methods used by others do not take
into account the effects related to internal grain structure and grain
surface roughness (see e.g. Henning and Stognienko 1993, Wolff  et al. 1994 and
Saija et al. 2001).

It must be emphasized that in the present study, using DDA for the composite grains, 
we have systematically studied the effects of inclusions and porosities and fit the 
IR emission from the circumstellar dust in the spectral range 5-25$\mu m$.
Further, with our composite grain model, we fit the average observed IRAS-LRS
emission curve obtained for circumstellar dust around several M-type stars (Whittet, 2003)
and two other individual IRAS stars.

{\tiny

\begin{table*}
\caption{Comparison of our model/results with available model/results from other workers.}
\begin{center}
\begin{tabular}{|l|c|c|c|c|}
\hline
            & Computational & Particle Shape & Inclusions/size        & Specifications of      \\
            & method        &                & and composition/       & model/method           \\
            &               &                & material               &                        \\
\hline
Our work    & DDA           & Oblate         & Inclusion size in terms & All composite grain\\
(2010)      &               &                & of number of dipoles      & models with volume \\
            &               &                & across the diameter      &fraction of inclusions\\
            &               &                & Inclusion compositions &viz. 10\%, 20\% and 30\%\\
            &               &                & are Graphite/ice \&    & were studied       \\
            &               &                & vacuum (for porosity)  &                    \\
\hline
Lee \&      & Rayleigh      & Oblate         & Ice, Size: smaller     & Core-mantle grains \\
Draine (1985) & approximation &              & than wavelength        &                        \\
\hline
Jones (1988) & EMA\tablefootmark{a} & Spheres        & Voids to model & Porosity 25\% and 50\%\\
            & Hollow spheres        &                & porous grains           &                      \\
\hline
Greenberg \&& EMA\tablefootmark{a} & Spheres        & Organic refractory  & Inclusion of CHON\\
Hage (1990) &               &                & material, Ice \&  & particles to model  \\
            &               &                & voids             & cometary grains             \\
\hline
Ossenkopf   & EMA\tablefootmark{a} & Ellipsoids     & Al2O3, MgO, MgS & 10\% inclusion of Al2O3,
\\
et al. (1992) &             &                & Fe3O4, Fe2O3, \&  & MgO and MgS                            \\
            &               &                & amorphous carbon  &                             \\
\hline
Henning \&  & DDA           & Oblate and Prolate  & Inclusion size smaller & Two fractions of  \\
Stognienko (1993) &  & The observed polarization  & than wavelength and    & inclusions        \\
            &         &  across the 10$\mu m$ & inclusion composition  & Fsilicates/Fgraphites \\
            &             &  feature led them     & consists of graphite,  & viz. 1.3 \& 0.8   \\
            &             & to conclude that the  & ice, voids \&           &                \\
            &             & silicate particles    & amorphous carbon        &                \\
            &             & are oblates           &                           &                \\
            &             & rather than prolates  &                           &                \\
\hline
O'Donnell   & EMA\tablefootmark{a} & Oblates and prolates  & Inclusion composition & Core-mantle particles with \\
(1994)      &        & They found that   & consists of & silicate-amorphous ice \\
            &        & prolate grains shift & amorphous carbon, & and amorphous carbon\\
            &        & the 10$\mu m$ \& 18$\mu m$  & glassy carbon,  &         \\
            &        & features too so as  & tholins \& voids            &         \\
            &        & to be consistent     &                                  &         \\
            &        & with the observed data    &                &             \\
\hline
Min et al.  & DDA     & Gaussian Random Spheres & Inclusion consists of & Silicate grains       \\
(2007)      &         & Gaussian Random         & distribution of     & with Mg component       \\
            &         & Field particles         & amorphous carbon,   & $>$ 0.9                 \\
            &         & hollow spheres          & amorphous silicate  &                         \\
\hline
Voshchinnikov \&  & Layered\tablefootmark{a} & Spherical grains & Inclusion composition & Volume fraction of inclusions  \\
Henning (2008) & spheres &               & consists of amorphous  & range from 0.2              \\
            &         &                  & carbon and amorphous   & to 0.9                      \\
            &         &                  & silicates   &                             \\
\hline
Li et al.   & Multi-layered\tablefootmark{a}  & Concentric  & Inclusion composition  & Mass ratio of amorphous \\
(2008) & sphere model   & spherical layers & consists of amorphous & carbon v/s amorphous silicates \\
       &                &                  & carbon \& porous/voids & to be $\sim$ 0.7          \\
\hline
\end{tabular}
\tablefoot{ 
\tablefoottext{a}{It is a mixture of two materials, there are no separate inclusions,
size is not applicable.}
}
\end{center}
\end{table*}
}

{\tiny

\begin{table*}
\caption{Comparison continued from previous table.}
\begin{center}
\begin{tabular}{|l|c|c|c|c|}
\hline
 & Characteristics of  & Characteristics of  & Ratio(R)  & Observational  \\
 & 10$\mu m$ feature  & 18$\mu m$ feature & 18$\mu m$/10$\mu m$ & characteristics \\
\hline
Our work  & Shifts shortwards with & No shift in this& R Decreases with & Dust temperatures 200-350K \\
(2010)    & graphite inclusions & feature is indicated & volume fractions of & are derived from composite \\
          &  \& no broadening of &           & graphite; increases    & grain models fit the \\
          & this feature is      &           & with porosity \&  &  observed IRAS-LRS curve\\
          & indicated            &           & varies between 0.2-0.6 & as well as for 2 stars \\
          &                      &           &                       & selected in this paper\\
\hline
Lee \&    & No shift is indicated & Not studied   & R not determined & Core-mantle grain model fit \\
Draine (1985) &                 &                 &                & well with the observed\\
          &                     &                 &                & IR emission from BN object\\
\hline
Jones     & Feature is     & Feature is  & R not determined & No comparison was made  \\
(1988)    & enhanced       & enhanced    &                &                           \\
\hline
Greenberg & Grains with silicate cores & This feature & R not determined & Pure silicate does not fit\\
\& Hage (1990) & \& refractory organic & was not &                & the observed emission in  \\
          & materials as mantles & studied       &           & comets but porous grains  \\
          & with $\sim$60-80\% porosity    &      &                & and CHON particles fit the\\
          & fit the 10$\mu m$ feature&            &                & observed IR emission from \\
          & observed in comets   &                &                & cometary dust             \\
\hline
Ossenkopf & Feature shifts shortwards  & Did not find  & R$\sim$0.3-0.5 & Grain models based on optical \\
et al. (1992) & ($\sim 0.3\mu m$) \& broadens & any shift in  & for most inclusions  & constants of \\
          & (FWHM 2-2.7$\mu m$)  & this feature & \& is enhanced to  & astronomical Si (Draine 1985)\\
          & with Al2O3 inclusions &        & 0.68 for MgO      & do not fit the observed\\
          &                       &        &                & observed IR emission from \\
          &                       &        &                &  circumstellar dust\\
\hline
Henning \& &  No appreciable shift& Feature shifts& R$\sim$0.41-0.47 & Porous and composite grains \\
Stognienko (1993) & with inclusions  & shortwards and&             & do not fit the observed  \\
          & indicated             & broadens with &                & IR emission from AFGL 2591 \\
          &                       & amorphous carbon&              & or BN objects       \\
\hline
O'Donnell & No shift in this      & No shift in   & R=0.51 for prolates & The silicate grains which give rise    \\
(1994)    & feature with          & this feature  & R=0.42 for oblates & to the 10$\mu m$ \& 18$\mu m$ features do not      \\
          & inclusions indicated  & with inclusions& Observed R value & posses any coatings of     \\
          &                       & is indicated     & is between 0.3-0.6 & amorphous or glassy carbon  \\
\hline
Min et al. & Feature shifts shortwards  & No shift in  & R not determined & GRF, DHS and irregularly \\
(2007)    & ($\sim 0.5\mu m$) \& broadens   & this feature  &  & shaped particles fit \\
          & (FWHM 2-2.5$\mu m$)             &               &  & the interstellar spectrum\\
          &                                 &               &  & in the region 5-25$\mu m$\\
\hline
Voshchinnikov & The strength of this & This feature  & R not determined & Porous and fluffy grain\\
\& Henning (2008) & feature decreases with  & was not&             & models fit the observed  \\
          & the porosity and the  & studied       &                & 10$\mu m$ feature in \\
          & feature also broadens &               &                & T-Tauri \& Herbig Ae/Be stars    \\
          & (FWHM 2.1-2.7$\mu m$) &               &                &                 \\
\hline
Li et al. & This feature shifts to longer & This feature  & R not determined & Observed 10$\mu m$ feature  \\
(2008)    & wavelengths ($\sim 10.6\mu m$) & was not    &     & can be explained in\\
          & \& the profile broadens        & studied    &  & terms of porous composite  \\
          &(FWHM 2.1-2.8$\mu m$)            &                       & & dust consisting of\\
          & with porosity     &                       & & amorphous Si and C \& vacuum\\
\hline
\end{tabular}
\end{center}
\end{table*}

}

\section{Summary and Conclusions}

We have used the discrete dipole approximation (DDA) to calculate
the absorption efficiency for the composite spheroidal grains and studied the
variation of absorption efficiency with the volume fractions of the inclusions
in the wavelength region of 5.0-25.0$\mu m$.
These results clearly show the variation in the absorption efficiency for the composite
grains with the volume fractions of the inclusions as well as
with porosity. The results on the composite
grains with graphite as inclusions show a shift towards shorter wavelength for
the peak absorption feature at 10$\mu m$ with the volume fraction of the inclusions.
However, these composite grain models did not show any
shift in the 18$\mu m$ peak with the variation in the inclusions or porosities.
Henning and Stognienko did not find any shift in either 10$\mu m$ or 18$\mu m$ feature for the
oblate composite spheroid containing silicates and graphites.
For the porous silicate grains, we did not find any shift in 10$\mu m$ or 18$\mu m$ feature;
whereas Li et al. (2008) found the shift towards longer wavelength in the 10$\mu m$ feature
for the porous silicate grains.
Ossenkopf et al. (1992) found shift towards shorter wavelength in the 10$\mu m$ feature
for the composite silicate grains with carbon inclusions.
The composite grain models presented in this study did not show any broadening in the
10$\mu m$ and 18$\mu m$ features. In the Table 6 \& 7 and Section 4, we compare and summarize 
all the results.

The dust temperatures between 200-350K derived from the composite grain models
fit with the observed IRAS-LRS curve and comparable with the dust temperature range 200-400K as
suggested by Voshchinnikov \& Henning (2008). The ratio R=Flux(18$\mu$)/Flux(10$\mu$) obtained
from the composite grain model varies from 0.2 to 0.6 and compares well with that 
derived from the observed IRAS-LRS curves for the circumstellar dust viz.
Little-Marenin and Little (1990); Ossenkopf et al. (1992) and Volk \& Kwok (1988).

It must be noted here that the composite grain models considered in the
present study are not unique.
However, these results on composite grains clearly indicate that the 
silicate features at 10$\mu m$ and 18$\mu m$ vary with the volume fraction of
inclusions and porosities. We also note that the results based on DDA and EMA calculations 
for the composite grains do not agree. The composite grain models presented in this 
paper need to be compared with a larger sample of stars with circumstellar
dust (Little-Marenin \& Little, 1990, Simpson 1991, 
Ossenkopf et al. 1992 and Henning \& Stognienko, 1993).

\begin{acknowledgements}
DBV and RG thank ISRO-RESPOND for the grant (N0. ISRO/RES/2/345/
2007-08), under which this study has been carried out.
Authors thank the referee for constructive suggestions.
\end{acknowledgements}

\end{document}